\documentclass[%
 reprint,
superscriptaddress,
 amsmath,amssymb,
 aps,
]{revtex4-2}

\usepackage{graphicx}
\usepackage{dcolumn}
\usepackage{bm}

\begin{document}

\preprint{APS/123-QED}

\title{Phase Transition in Extended Thermodynamics Triggers Sub-shocks}

\author{Candi Zheng}%
 \email{Corresponding author, czhengac@connect.ust.hk}
\affiliation{%
Department of Mechanics and Aerospace Engineering, Southern University of Science and Technology, Xueyuan Rd 1088, Shenzhen, China 
}%
\affiliation{%
  Department of Mathematics,
  Hong Kong University of Science and Technology, Clear Water Bay, Hong Kong SAR, China
}%
 \author{Yang Wang}%
 \email{yangwang@ust.hk}
 \affiliation{%
  Department of Mathematics,
  Hong Kong University of Science and Technology, Clear Water Bay, Hong Kong SAR, China
}%
  \author{Shiyi Chen}%
 \email{ chensy@sustech.edu.cn}
 \affiliation{%
Eastern Institute for Advanced Study, Eastern Institute of Technology, Ningbo, Zhejiang 315200, P. R. China.
}%
\affiliation{%
Department of Mechanics and Aerospace Engineering, Southern University of Science and Technology, Xueyuan Rd 1088, Shenzhen, China 
}%
\date{\today}

\begin{abstract}
Extended thermodynamics commonly uses polynomial moments to model non-equilibrium transportation, but faces a crisis due to sub-shocks, which are anomalous discontinuities in gas properties when predicting shock waves. The cause of sub-shocks is still unclear, challenging the validity of extended thermodynamics. This paper reveals, for the first time, that sub-shocks arise from intrinsic limitations of polynomials leading to a discontinuous phase transition. Therefore extended thermodynamics necessitates alternative moments beyond polynomials to avoid sub-shocks.
\end{abstract}

\maketitle


\textbf{Introduction} Non-equilibrium transportation process, such as shock waves in gas, play a crucial role in understanding various phenomena in aerospace engineering \cite{Chen1998LATTICEBM, alma991005956889703412}, astrophysics \cite{blandford1987particle, AbarzhiS.I.2013Tmab} and condensed matter \cite{CamiolaVitoDario2020CTiL}. Traditional approaches like the Navier-Stokes-Fourier (NSF) equations \cite{agarwal2001beyond, GarcaColn2008BeyondTN}, which are based on the conservation of mass, momentum, and energy, often struggle to accurately describe non-equilibrium processes. In contrast, extended thermodynamics provides a significantly more computationally efficient framework for simulating non-equilibrium dynamics governed by the Boltzmann transport equation \cite{Cercignani1988,alma991012694041603412,TorrilhonManuel2016MNGF,McDonaldJames2013Armc,Schaerer2017EfficientAA}, compared to other methods such as direct simulation Monte Carlo (DSMC) \cite{Bird1994} and discrete velocity methods (DVM) \cite{DVMbook}. Extended thermodynamics incorporates additional polynomial moments of a particle's probability density distribution \cite{alma991012694041603412}, offering a detailed and accurate representation of non-equilibrium transportation of gas molecules or electrons \cite{TorrilhonManuel2016MNGF, CamiolaVitoDario2020CTiL}.


However, a critical problem in extended thermodynamics, present since the 1950s, is the occurrence of unrealistic sub-shocks in predicted gas shock wave profiles \cite{Grad1952ThePO}. Sub-shocks are discontinuous changes in gas properties within shock wave profiles, conflicting with experimental observations and raising a crisis for extended thermodynamics, questioning the theoretical soundness of its current formulation \cite{Israel, alma991012694041603412, BoillatGuy1998Otss}. Resolving the sub-shock issue is crucial for justifying extended thermodynamics as a valid model for non-equilibrium gas dynamics.


Understanding the sub-shock formation is essential for resolving the sub-shock issue. Previous research on sub-shock formation has attempted to identify potential causes using linearization, spectral analysis, and entropy inequalities \cite{BoillatGuy1998Otss, RuggeriTommaso2022Acco, WeissW1995Cssi,TaniguchiShigeru2018Otsf, BisiMarzia2016SfiG}. These studies have revealed that sub-shocks may appear when the shock velocity meets any characteristic velocity of the hyperbolic system, but whether they can only form at the maximum characteristic velocity remains controversial \cite{BoillatGuy1998Otss, RuggeriTommaso2022Acco, TaniguchiShigeru2018Otsf}. Furthermore, the Holway-Weiss debate, which questions whether the restricted convergence of moment expansions causes sub-shocks, has yet to be resolved \cite{Holway1964, WeissW1995Cssi,CaiZhenning2019OtHd}. Despite various efforts to understand sub-shock formation, current sub-shock formation theories have not been able to identify the cause of sub-shocks conclusively.

\begin{figure}
\includegraphics[width=3.5in]{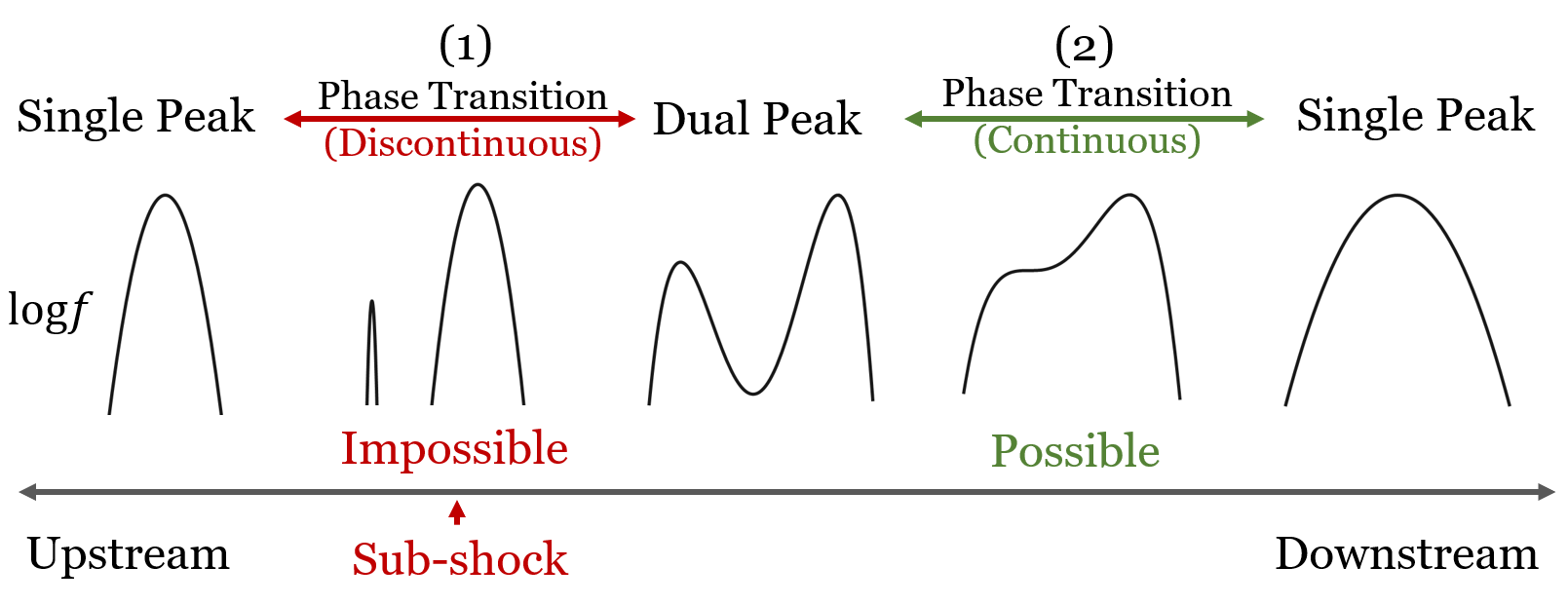}
\caption{\label{fig:PhaseChanges} Illustration of the shape change of the particle velocity distributions $f$ across the shock wave, highlighting two distinct phase transitions. (1) Phase transition from a single peak to a dual peak, where the 35 polynomial moment system fails to describe the emergence of an additional narrow peak, leading to a discontinuous change and the formation of a sub-shock. (2) Phase transition from a dual peak to a single peak, characterized by a smooth merging of the two peaks, ensuring a continuous transition without the occurrence of a sub-shock.}
\end{figure}

This paper identifies the cause of sub-shocks in extended thermodynamics for the first time. Our analysis reveals that sub-shocks are caused by a previously unrecognized phase transition. This phase transition is a discontinuous change in the shape of particle velocity distributions $f$, shifting from single-peaked to dual-peaked profiles as the gas passes through the shock front (FIG.\ref{fig:PhaseChanges}). The discontinuity stems from polynomials' intrinsic limitations in describing peak emergence continuously. Our findings elucidate the formation of sub-shocks as phase transitions, attributing the sub-shock crisis to the limitation of polynomials rather than the framework of extended thermodynamics. This essential step towards resolving the sub-shock crisis highlights the need to explore alternative moments for continuous sub-shock handling.

\textbf{Method} This paper focuses on normal shock waves in gases, which are fast-propagating disturbances causing sudden and intense changes in properties such as density, velocity, and temperature. The undisturbed gas flows toward the shock wave (upstream) and then passes through it, undergoing intense changes (downstream). The upstream and downstream properties are connected by the Rankine-Hugoniot condition through the Mach number \cite{Landau1987}, given by $M = v_{-}\sqrt{m/\gamma k_B T_{-}}$, where $v_{-}$ denotes upstream flow velocity, $m$ represents gas molecules' mass, $\gamma$ is the heat capacity ratio, and $k_B$ is the Boltzmann constant. The sudden and intense nature of shock waves makes them strongly non-equilibrium phenomena.

As traditional NSF equations struggle to accurately represent strong non-equilibrium dynamics in shock waves, extended thermodynamics offers a superior solution. By considering higher-order moments of the distribution function $f(\mathbf{r}, \mathbf{u}, t)$—indicating the probability of locating particles with velocity $\mathbf{u}$ at position $\mathbf{r}$ and time $t$—extended thermodynamics captures the intricate non-equilibrium processes within the shock front. This results in a more precise portrayal of shock wave dynamics.



High-order moments are crucial components in extended thermodynamics. They are defined as:
\begin{equation} \label{moments integration}
M_n(\mathbf{r}, t) = \int \phi_n(\mathbf{u}) f(\mathbf{r}, \mathbf{u}, t) d\mathbf{u},
\end{equation}
where $n$ denotes the order of the moment, and $\phi_n(\mathbf{u})$ are high-order polynomials of particle velocity $\mathbf{u}$ depending on the choice of moments in the model.

The key assumption of extended thermodynamics is that a finite set of prescribed moments can completely model the distribution function $f(\mathbf{r}, \mathbf{u}, t)$. It allows us to model the distribution with the exponential family model \cite{MAL-001ExpFamily} expressed as:
\begin{equation} \label{exponential family model}
\begin{split}
    \log f(\mathbf{r}, \mathbf{u}, t) &= \sum_{n = 0}^{N} \lambda_n(\mathbf{r}, t) \cdot \phi_n(\mathbf{u}) ,
\end{split}
\end{equation}
where $N$ is the highest order moment considered, $\phi_n(\mathbf{u})$ are prescribed polynomials, and $\lambda_n(\mathbf{r}, t)$ are the Lagrange multipliers. This model is equivalent to the maximum entropy moment method \cite{levermore1996moment}. Additionally, the Grad moment method \cite{Grad1949} is a special case of this method when the higher-order moments are close to their equilibrium value.

\begin{figure}
\includegraphics[width=3.4in]{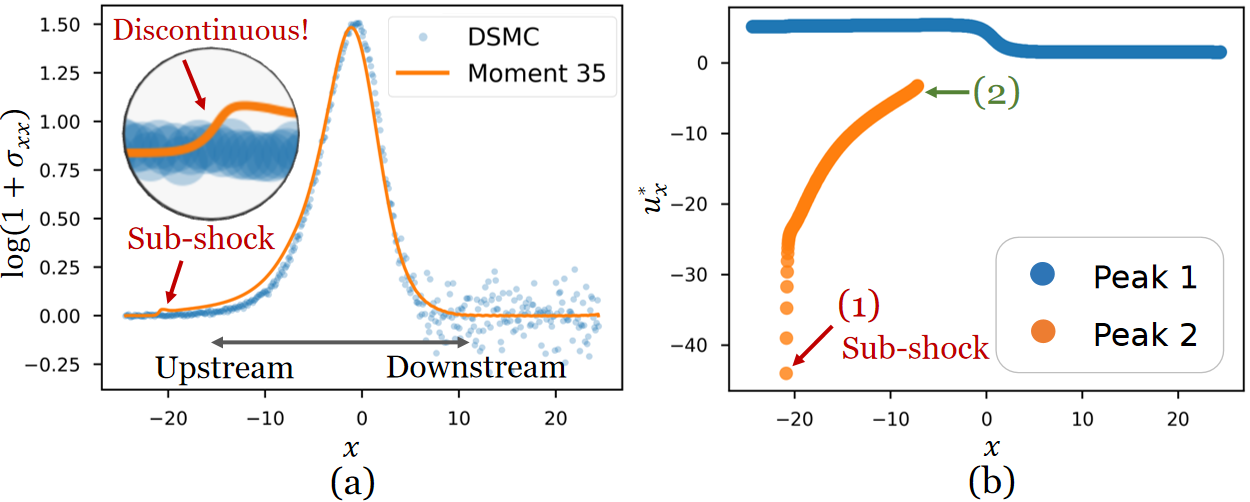}
\caption{\label{fig:Subshocks} (a) Stress profile comparison for a Mach 4 normal shock wave: Extended thermodynamics using the 35 polynomial moment system vs. DSMC method, illustrating the discontinuous sub-shock in the extended thermodynamics result due to the method's limitations; (b) Local maximums $u^*$ of the distribution function $f(\mathbf{u},x)$, highlighting the emergence of a second peak (peak 2) by phase transition (1) in Fig.\ref{fig:PhaseChanges} and its disappearance by phase transition (2). The emergence of the second peak in (b) at the same position as the sub-shock in (a) triggers the sub-shock.}
\end{figure}

The backbone of extended thermodynamics is the moment equations, which are conservation laws governing the time evolution of moments $M_n$ and Lagrangian multipliers $\lambda_n$. Derived by substituting the exponential family model into the Boltzmann equation, these equations can be expressed as
\begin{equation}
\partial_t M_n + \sum_{\alpha \in \{x,y,z\}} \partial_\alpha F_{n,\alpha} = \mathbf{S}_n,
\end{equation}
where $F_{n,\alpha}$ denotes the flux of the $n$-th moment in the $\alpha$ direction, while $\mathbf{S}_n$ represents a source term accounting for physical processes such as collisions, radiative transport, and chemical reactions. We solve these moment equations for a Mach 4 normal shock wave  numerically using the Lax-Friedrich method, with details on flux and source terms, boundary conditions, and numerical schemes provided in \cite{zheng2023stabilizing}. The moment equations produce a shock wave that agrees with the DSMC reference profile in Fig.\ref{fig:Subshocks}(a), except for a serious defect: a discontinuous sub-shock at the shock front that violates the continuity of shock wave profile.

The discontinuous sub-shock poses a significant challenge to the theoretical soundness of extended thermodynamics \cite{Israel, alma991012694041603412, BoillatGuy1998Otss}, necessitating a thorough investigation. Traditional analysis of sub-shock, which examines local characteristic speeds and entropy \cite{BoillatGuy1998Otss, RuggeriTommaso2022Acco, WeissW1995Cssi,TaniguchiShigeru2018Otsf, BisiMarzia2016SfiG}, fails to explain the formation of sub-shock physically due to limited information on the shape of distribution $f$. To address this, we explore a new analysis based on saddle point approximation.

We direct our analysis towards understanding the shape of the distribution function, specifically its peaks (local maxims), through the use of the saddle point approximation. The partition function denoted as $Z(\boldsymbol{\lambda})$ is the moment of the exponential family model \eqref{exponential family model} when $\phi(\mathbf{u}) = 1$. The saddle point approximation \cite{Kardar2007StatisticalPOC49} enables the estimation of the partition function by summing the contributions from multiple maximums in the distribution function $f$: 
\begin{equation} \label{saddle approx}
Z_{saddle}(\boldsymbol{\lambda}) \approx \sum_{k=1}^{p} \sqrt{\frac{ (2\pi)^3 }{ \left|\det H_k(\boldsymbol{\lambda}) \right| }}\exp{\left(\sum_{n = 0}^{N} \lambda_n \cdot \phi_n(\mathbf{u}^*_k) \right)}, 
\end{equation} 
where $p$ represents the number of local maxima and $H_k(\boldsymbol{\lambda})$ is the Hessian matrix of $\log f(\mathbf{r}, \mathbf{u}, t)$ evaluated at the $k$-th local maxima $\mathbf{u}^*_k$. The distribution's moments, $M_n(\mathbf{r}, t)$, are subsequently calculated as: 
\begin{equation} M_n(\mathbf{r}, t) \approx \frac{\partial Z_{saddle}(\boldsymbol{\lambda})}{\partial \lambda_n}. 
\end{equation}
The saddle point approximation allows us to determine the number of peaks and their contribution to moments. A phase transition occurs when the number of peaks changes as the gas traverses the shock front, resulting in sub-shocks in extended thermodynamics.

\begin{figure}
\includegraphics[width=3.4in]{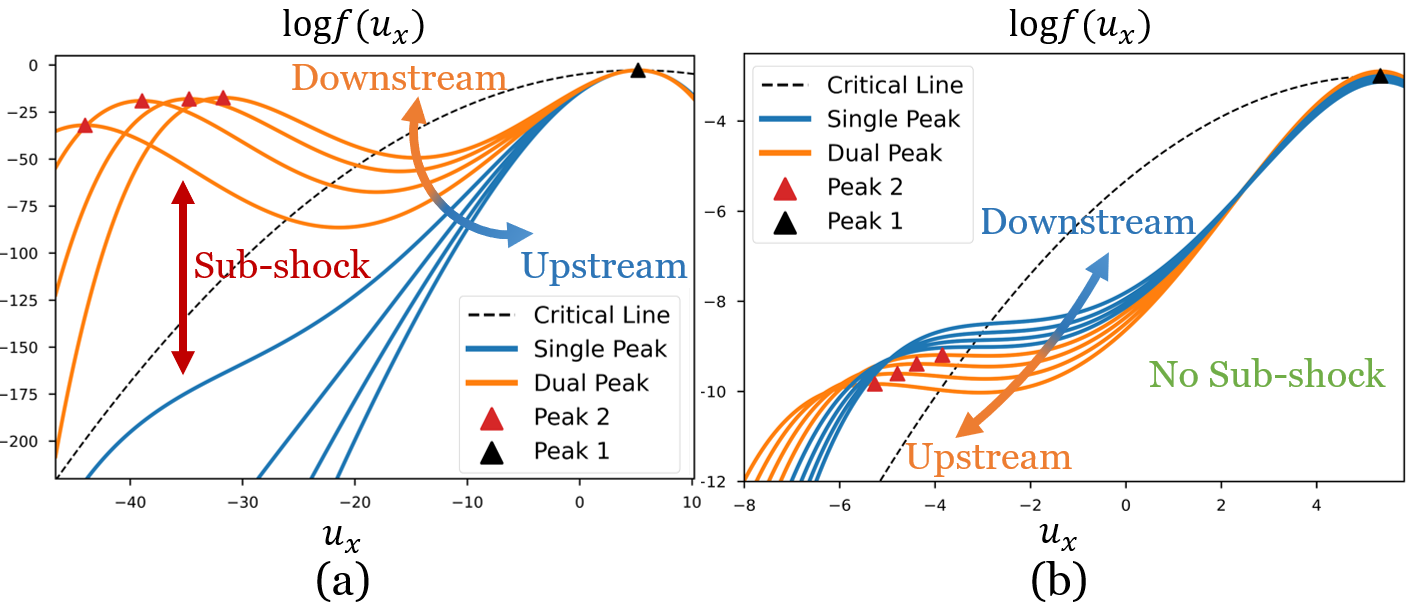}
\caption{\label{fig:PhaseD} (a) The discontinuous emergence (sub-shock) of the second peak near phase transition (1) in FIG.\ref{fig:PhaseChanges}; (b) the continuous disappearance of the second peak near phase transition (2) in FIG.\ref{fig:PhaseChanges}. Both diagrams (a) and (b) plot log-likelihoods $\log f(u_x)$ against particle velocity $u_x$ with local maximums highlighted as peaks 1 and 2. A critical line separates single- and dual-peak phases, representing peak 2 positions with zero curvature. Diagrams (a) depicts a sudden phase transition from single- to dual-peak, with a discontinuous emergence of peak 2 causing the sub-shock. Diagrams (b) reveals a continuous transition from dual- to single-peak as peak 2 approaches the critical line and vanishes with no sub-shock. These plots demonstrate that the sub-shock arises from the discontinuous emergence of peak 2 during the phase transition.}
\end{figure}

We apply the saddle point approximation to the 35 moment system \cite{Schaerer2017EfficientAA, Schaerer2017The3S} as it is a typical scenario admitting two peaks, which is consistent with common assumptions in the Mott-Smith model \cite{MottSmith1951TheSO}. For one-dimensional flow, this moment system is defined by polynomial statistics $\phi_n(\mathbf{u})$ as
\begin{equation}
\label{Sufficient 35}
\begin{split}
\{\phi_i(\bold{u})\} &= \{1, u_x, u_x^2, u_x^3, u_x^4, u_r^2, u_r^4, u_x u_r^2, u_x^2 u_r^2\},
\end{split}
\end{equation}
where $u_r = \sqrt{u_y^2 + u_z^2}$. In this system, the two local maxims of the distribution function $f$ can be found in the form $\mathbf{u}_{1,2}^* = \{u_{1,2}^*,0,0\}$, where $u_{1,2}^*$ corresponds to the two local maxims of the polynomial \begin{equation}\label{35 poly}
\log f(u_x) = \lambda_0 + \lambda_1 u_x + \lambda_2 u_x^2 + \lambda_3 u_x^3 + \lambda_4 u_x^4,
\end{equation} 
in which $\lambda$ are the Lagrange multiplies in \eqref{exponential family model}. Our analysis of the 35-moment system is applicable to any higher-order polynomial moment systems, which, as discussed in the discussion section, tend to exhibit more sub-shocks and poorer peak resolution.

\begin{figure}
\includegraphics[width=2.6in]{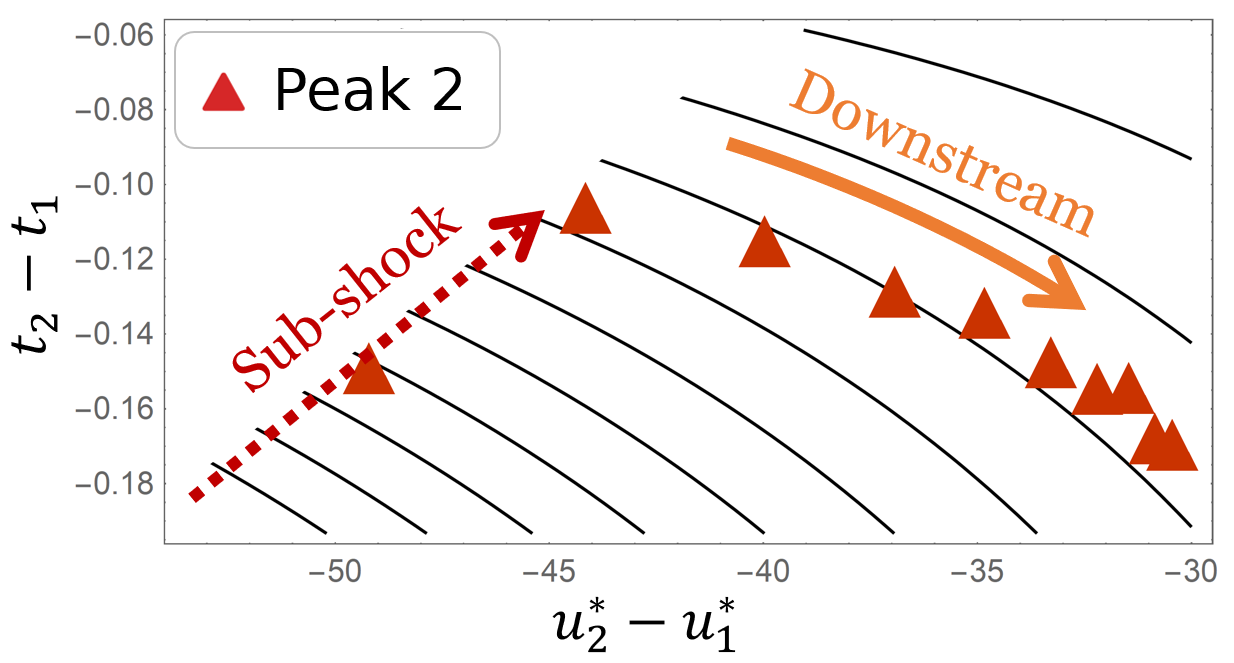}
\caption{\label{fig:Curv} Relative location and curvature of Peak 2 downstream of the sub-shock, with solid lines indicating constant stress ( contour lines of \eqref{sigma equation}). The figure demonstrates that Peak 2 arises with the sub-shock as a narrow peak and then aligns with the constant stress lines, exhibiting a decreasing curvature in the downstream direction. This decreasing curvature confirms that Peak 2, arising as a narrow peak, broadens in the downstream direction.}
\end{figure}

Sub-shocks arise due to the inability of polynomial \eqref{35 poly} to describe two independent peaks results. The polynomial has only five free $\lambda$ parameters, inadequate for describing the two local maxima, $u_{1,2}^*$, which require at least six degrees of freedom. According to the saddle point approximation, each local maximum needs three degrees of freedom: the maximum value, the vanishing first derivative, and the curvature required in the Hessian. Consequently, the maxima are intertwined, with their maximum values, locations, and curvatures connected by the shape equation:
\begin{equation} \label{shape equation}
\log f_2 - \log f_1=\frac{1}{12} (t_2-t_1) (u_{2}^*-u_{1}^*)^2,
\end{equation}
in which $u_{1,2}^*$ are the local maximums of \eqref{35 poly}, $\log f_{1,2} = \log f(u_{1,2}^*)$ are the maximum values, and $t_{1,2} = -\log f''(u_{1,2}^*)$ are the curvature at each local maximum. This connection results in a constraint on the relative amplitude and curvature. When the peak $2$ at $u_{2}^*$ has a smaller amplitude than the peak $1$ at $u_{1}^*$, it must also have a smaller curvature
\begin{equation} \label{curv inequation}
t_2 < t_1
\end{equation}
because the sign of $\log f_2 - \log f_1$ must be the same as $(t_2-t_1)$.
Therefore, peak $2$ cannot be narrower than the peak $1$, precluding a smooth transition from single-peak to dual-peak profiles initiated by a narrow, low-amplitude peak 2, as shown in phase transition (1) in FIG.\ref{fig:PhaseChanges}. This results in a discontinuous phase change, producing sub-shocks.

To verify that phase transition (1) is initiated by a narrow and low-amplitude second peak, we examine the distribution $f$'s behavior near the sub-shock. Assuming isotropic peaks with Hessian $H_k = -t_k \mathbf{I}$, where $\mathbf{I}$ is the identity matrix, we analyze the stress near the sub-shock using the saddle point approximation. Upstream of the sub-shock, the single-peak distribution function $f$ yields zero stress, confirmed by Figure 2(a). Downstream, however, $f$ possesses two peaks, with the newly emerged peak 2 at $u_2^*$ contributing most to the stress, as described by
\begin{equation} \label{sigma equation}
\sigma_{xx} \propto t_2^{-3/2}(u_1^*-u_2^*)^2 e^{\frac{1}{12} (t_2-t_1) (u_1^*-u_2^*)^2},
\end{equation}
in which the location $u_2^*$ of peak 2 changes significantly and rapidly immediately downstream of the sub-shock without significant disturbance in stress, as shown in FIG.\ref{fig:Subshocks}. Consequently, $u_2^*$ and its curvature $t_2$ lie on the constant $\sigma_{xx}$ contour line plotted in FIG.\ref{fig:Curv}. This results in the peak 2 widening downstream from the sub-shock as $u_2^*$ increases and $t_2$ decreases, confirming the occurrence of phase transition (1) in FIG.\ref{fig:PhaseChanges}, where a narrow second peak appears and gradually widens.

\textbf{Results} Extended thermodynamics with the 35-moment system predicts two distinct phase transitions across a Mach 4 normal shock wave profile, as demonstrated by phase transitions (1) and (2) in FIG.\ref{fig:PhaseChanges}. Evidence for these phase transitions is provided by FIG.\ref{fig:Subshocks}(b), which shows the emergence and subsequent disappearance of a second peak (peak 2). Further analysis of the log-likelihood near the phase transitions in FIG.\ref{fig:PhaseD} confirms the discontinuous emergence and smooth disappearance of this second peak. It is the discontinuous emergence of the second peak that leads to the formation of a sub-shock, which is confirmed by its alignment with the occurrence of the sub-shocks in Figure 2(a) and (b). The sub-shock is an unavoidable consequence of the 35-moment system, as demonstrated in the following argument through a proof by contradiction.

To establish the inevitability of the sub-shock, we first list the two approaches to make the emergence of the second peak continuous: (I) a flat (zero curvature) second peak emerges with a very small amplitude, with both amplitude and curvature increasing in the downstream direction, and (II) a spike-like (infinitely large curvature) second peak emerges with finite amplitude smaller than the first peak. These approaches ensure that the contribution of the second peak to the system is small, which is proportional to second peak's amplitude and inversely proportional to its curvature.

Next, we show that both of these proposed methods are impossible. Approach (I) is untenable because FIG.\ref{fig:Curv} demonstrates that the curvature of the second peak decreases as we move in the downstream direction. It contradicts the increasing curvature required by approach (I). Approach (II) is also impossible because, according to \eqref{curv inequation}, the curvature of the second peak must be smaller than that of the first peak. Consequently, neither approach is feasible, so the second peak cannot emerge continuously, and the sub-shock cannot be avoided.

In summary, our results establish the existence of two phase transitions as the emergence of and disappearance of the second peak. It is the discontinuous emergence of the second peak causing the formation of a sub-shock, which is unavoidable due to the inherent limitation of the extended thermodynamics with the 35-moment system.

\textbf{Discussion}
This paper propose a novel explanation for the formation of sub-shock waves, challenging the prevailing understanding that attributes sub-shock occurrence to shock wave speeds exceeding all characteristic speeds \cite{WeissW1995Cssi}. Our findings indicate that shock wave speed is not the fundamental factor for sub-shock formation; rather, it serves as a prerequisite for the emergence of multimodal distributions exhibiting multiple peaks. Such distributions can describe high-speed shock waves far from equilibrium but do not necessarily result in sub-shock formation. The essential cause of sub-shock emergence lies in the inability of the polynomial moment system to accurately represent smooth transitions between different modes, leading to discontinuous transitions that give rise to sub-shocks.

Our analysis of the 35-moment system can be generalized to higher-order polynomial moment systems. Surprisingly, higher-order polynomials exacerbate the sub-shock issue and degrade peak resolution. This is because introducing higher-order polynomials into the system impairs its ability to characterize transitions between different peaks, contrary to expectations. According to the saddle point approximation, a polynomial of order $2n$ possesses $2n+1$ degrees of freedom, while its $n$ peaks require $3n$ degrees of freedom for full characterization. As a result, when $n>1$, full characterization of all peaks becomes unattainable due to $3n > 2n+1$, leading to intertwined peaks with a compromised ability to represent peak emergence and disappearance. Consequently, higher-order polynomials result in more intertwined peaks, thereby aggravating the sub-shock issue.

Our study observed sub-shocks with notably smaller amplitudes and more upstream positions compared to earlier studies \cite{Schaerer2017The3S, Schaerer2017EfficientAA}. This discrepancy is primarily attributed to differences in the integration domain for particle velocity ($u_x \in [a, b]$) when calculating moments using Eq. (1). Earlier studies employed a finite integration domain with a lower boundary $a$ that was insufficient to resolve the second peak fully. Consequently, a more significant sub-shock emerged when the the center of the second peak (FIG.\ref{fig:Subshocks}(b)) meets the lower integration boundary $u_x = a$, resulting in a more downstream sub-shock location compared to our results.

A possible way for mitigating the sub-shock issue involves incorporating hydrodynamic fluctuations into the extended thermodynamics framework. The current moment equation system, which is a first-order PDE, neglects hydrodynamic fluctuations that typically contribute to second-order dissipation terms. In three-dimensional space, fluctuations are known to significantly alter system behavior near phase transitions \cite{kardar_2007}. Consequently, accounting for hydrodynamic fluctuations may transform the discontinuous phase transitions causing sub-shocks into continuous transitions.

Another alternative approach to addressing the sub-shock issue involves substituting the polynomial model with more flexible methods, such as splines or neural networks. While polynomials excel at approximating local profiles near a single maximum, they struggle to accurately represent multiple maxima simultaneously, particularly in phase transitions where maxima emerge and vanish. Employing more flexible models like splines or neural networks could yield continuous descriptions of phase transitions, effectively eliminating sub-shocks.

In summary, we have identified the underlying cause of sub-shocks in extended thermodynamics, revealing their origin in the intrinsic limitations of polynomial moments to describe peak emergence continuously during phase transitions. This insight underscores the need to consider alternative moment systems capable of continuous representation of these transitions to address the sub-shock problem. Possible solutions include incorporating hydrodynamic fluctuations into the framework, or replacing polynomial models with more flexible methods like splines or neural networks.

\begin{acknowledgments}
We express our gratitude to Dr. Yuan Lan from the Hong Kong University of Science and Technology for her valuable feedback on the manuscript.
\end{acknowledgments}

\appendix

\nocite{*}

\bibliography{apssamp}

\end{document}